\documentclass[twocolumn,floatfix,amsmath,amssymb,prl,showpacs,superscriptaddress]{revtex4-1}
\usepackage{graphicx}

\usepackage{dcolumn}% Align table columns on decimal point
\usepackage{bm}% bold math
\usepackage{amsmath}
\begin{document}
\title{Transport signatures of pseudo-magnetic Landau levels in strained graphene ribbons}
\author{Diana A. Gradinar}
\email[Electronic mail: ]{d.cosma@lancaster.ac.uk}
\affiliation{Department of Physics, Lancaster University, LA1 4YB Lancaster, United Kingdom}
\author{Marcin Mucha-Kruczy\'{n}ski}
\affiliation{Department of Physics, Lancaster University, LA1 4YB Lancaster, United Kingdom}
\affiliation{Department of Physics, University of Bath, Claverton Down, Bath, BA2 7AY, United Kingdom}
\author{Henning  Schomerus}
\affiliation{Department of Physics, Lancaster University, LA1 4YB Lancaster, United Kingdom}
\author{Vladimir I. Fal'ko}
\affiliation{Department of Physics, Lancaster University, LA1 4YB Lancaster, United Kingdom}
\date{\today}
\begin{abstract}
In inhomogeneously strained graphene, low-energy electrons experience a valley-antisymmetric pseudo-magnetic field which leads to the formation of localized states at the edge between the valence and conduction bands, understood in terms of peculiar $n=0$ pseudo-magnetic Landau levels. Here we show that such states can manifest themselves as an isolated quadruplet of low-energy conductance resonances in a suspended stretched graphene ribbon, where clamping by the metallic contacts results in a strong inhomogeneity of strain near the ribbon ends.
\end{abstract}
\pacs{73.22.Pr, 62.20.-x, 71.70.Di}

\maketitle
Graphene~\cite{castro-neto1} is a one-atom-thick crystalline membrane~\cite{membrane} capable of withstanding reversible deformations of up to $10\%$~\cite{castro_neto}, which is important because of the peculiar way strain affects the electronic properties of this material. It is a common feature of all materials with several degenerate valleys in the band structure~\cite{koshelev} that the effect of their lattice deformations on electrons is equivalent to that of an effective gauge field~\cite{castro_neto,guinea,vozmediano}. In graphene, electrons near the Fermi level occupy states in the vicinity of a Dirac point at the edge between the valence and conduction bands, in one of two inequivalent valleys centered at the corners of the hexagonal Brillouin zone, $K$ and $K'$. Consequently, inhomogeneous strain in graphene influences electron motion in a manner similar to an effective pseudo-magnetic field, which has the opposite sign in the two valleys~\cite{suzura}. Recent scanning-tunneling experiments on graphene nanobubbles~\cite{crommie} have shown that even for relatively weak deformations such pseudo-magnetic fields can reach values equivalent to tens, and even hundreds of Tesla, resulting in the formation of a discrete spectrum of `Landau levels'~(LL), including the peculiar $n=0$ LL state pinned to the edge between the valence and conduction bands.

Here we show that such pseudo-magnetic LL states form in the contact regions of stretched graphene nanoribbons (GNR) and then give rise to a characteristic signature in the electronic transport: a quadruplet of low-energy conductance resonance, slightly split by the valley mixing and the tunnel coupling via evanescent modes in the middle part of the GNR.

\begin{figure*}[t!]
\includegraphics[width=2\columnwidth]{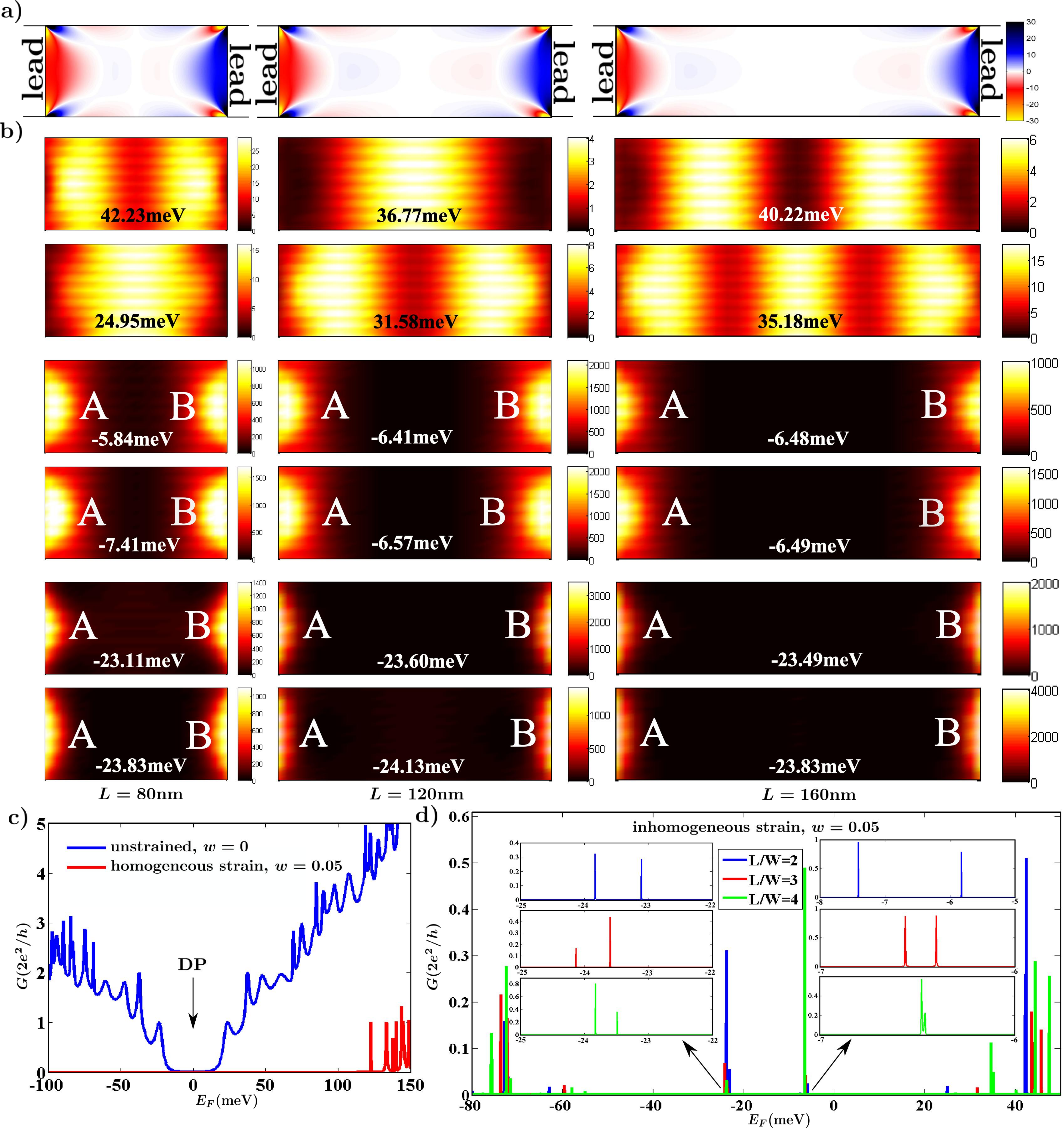}
\caption{\label{fig1} We consider the transport through strained suspended graphene nanoribbons (GNR) which are clamped at highly-doped contacts. Panel (a) shows the distribution of pseudo-magnetic fields $\mathcal{B}(\mathrm{T})$ for electrons in the $K$ valley for GNR with $W\simeq40$nm and aspect ratios $L/W=2$, $3$, and $4$. The inhomogeneous tensile strain in the middle of the nanoribbon is $w=0.05$. Panel (b) shows the spatial structure of electron wave amplitudes corresponding to several resonances identified in panel (d), which displays the zero-temperature conductance of the ribbons as a function of the Fermi energy. For comparison, panel (c) shows the conductance for the ribbon with $L/W=3$ and no strain~($w=0$, blue) or artificially imposed homogeneous strain~($w=0.05$, red).}
\end{figure*}

The considered device is comprised of a GNR that is clamped at the ends, and contacted by heavily doped unstrained graphitic leads. We choose the ribbon to have armchair side edges along the transport direction $x$, and set contacts with bulk electrodes along the $y$ direction. Such a ribbon can be described using the tight-binding Hamiltonian~\cite{castro-neto1}
\begin{align}
\mathcal{H}=\sum_i V_i c^{\dagger}_i c_i+\sum_{\left< ij\right>} \gamma_{ij}c^{\dagger}_i c_j,
\label{eqn-hamiltonian}
\end{align}
where $c_i$ is a fermionic annihilation operator acting on a site $i$ and $\left< ij \right>$ denote pairs of nearest neighbors. Compared to pristine monolayer flakes, the on-site potential $V_i$ is modulated by strain, which we take into account by $V_i=\frac{1}{2}\frac{\partial \epsilon_{c}}{\partial r}\mathrm{div} \boldsymbol{u}(\boldsymbol{r}_{i})$ where $\boldsymbol{u}=(u_x,u_y)$ is the displacement field of the membrane and $\epsilon_{c}$ is the on-site energy of electrons in a lattice with a given carbon-carbon bond length $r=1.42\AA$. The hopping matrix elements
\begin{equation}
\gamma_{ij}= \gamma_0 e^{\eta_0 (l_{ij}/r-1)}, \quad l_{ij}\simeq r(1+ \boldsymbol{n}_{ij} \cdot \boldsymbol{\hat{w}}\boldsymbol{n}_{ij}),
\label{eqn-hopping_renormalization}
\end{equation}
depend on the distance $l_{ij}$ between lattice sites, modified by the strain~\cite{castro_neto}. Here $\boldsymbol{\hat{w}}$ is the $2$x$2$ strain tensor $w_{\alpha\beta}=\frac{1}{2}(\partial_{\alpha} u_{\beta}+\partial_{\beta} u_{\alpha})$ with $\alpha,\beta=x$ or $y$, the parameter $\eta_0=\frac{\partial \mathrm{ln} \gamma_0}{\partial \mathrm{ln} r}\approx -3$ relates the change of the nearest neighbor coupling to the change of the bond length~\cite{john} ($\gamma_0\approx-3$eV), and $\boldsymbol{n}_{ij}=(1,0)$, $(-\frac{1}{2},\frac{\sqrt{3}}{2})$, and $(-\frac{1}{2},-\frac{\sqrt{3}}{2})$ are the unit vectors along carbon-carbon bonds in the honeycomb lattice. For a homogeneously strained armchair GNR, $w_{xx}=w$, $w_{yy}=-\sigma w$, and $w_{xy}=0$, where $\sigma=0.165$ is the Poisson ratio for graphite~\cite{proctor} and $w$ parameterizes tensile strain. The strain-induced asymmetry in the hoppings between neighboring carbon sites is equivalent to the effect of a valley-dependent vector potential
\begin{align}
e\boldsymbol{\mathcal{A}}=\xi\frac{\hbar \eta_0}{2 r} \left(
\begin{array}{c}
w_{xx} -w_{yy}  \\
-2w_{xy} \end{array} \right),\nonumber
\end{align}
where $\xi=\pm 1$ for valleys $K$ and $K'$. Clamping the ends of a suspended strained ribbon makes the strain distribution near the device contacts inhomogeneous~\cite{f_note}, and this results in the appearance of a pseudo-magnetic field $\boldsymbol{\mathcal{B}}=\mathrm{rot}\boldsymbol{\mathcal{A}}$ in the vicinity of the contacts \cite{suzura,marcin}. This is illustrated in Fig.~\ref{fig1}(a), where we applied linear elasticity theory~\cite{marcin,timoshenko} to three ribbons of width $W\simeq40$nm and aspect ratio $L/W=2$, $3$, and $4$, with $w=0.05$ in their middle parts. The pseudo-magnetic field is the largest positive~(blue) or negative~(red) near the contacts at the right and left ends, and is small in the middle of the ribbon.

\begin{figure}[t!]
\includegraphics[width=\columnwidth]{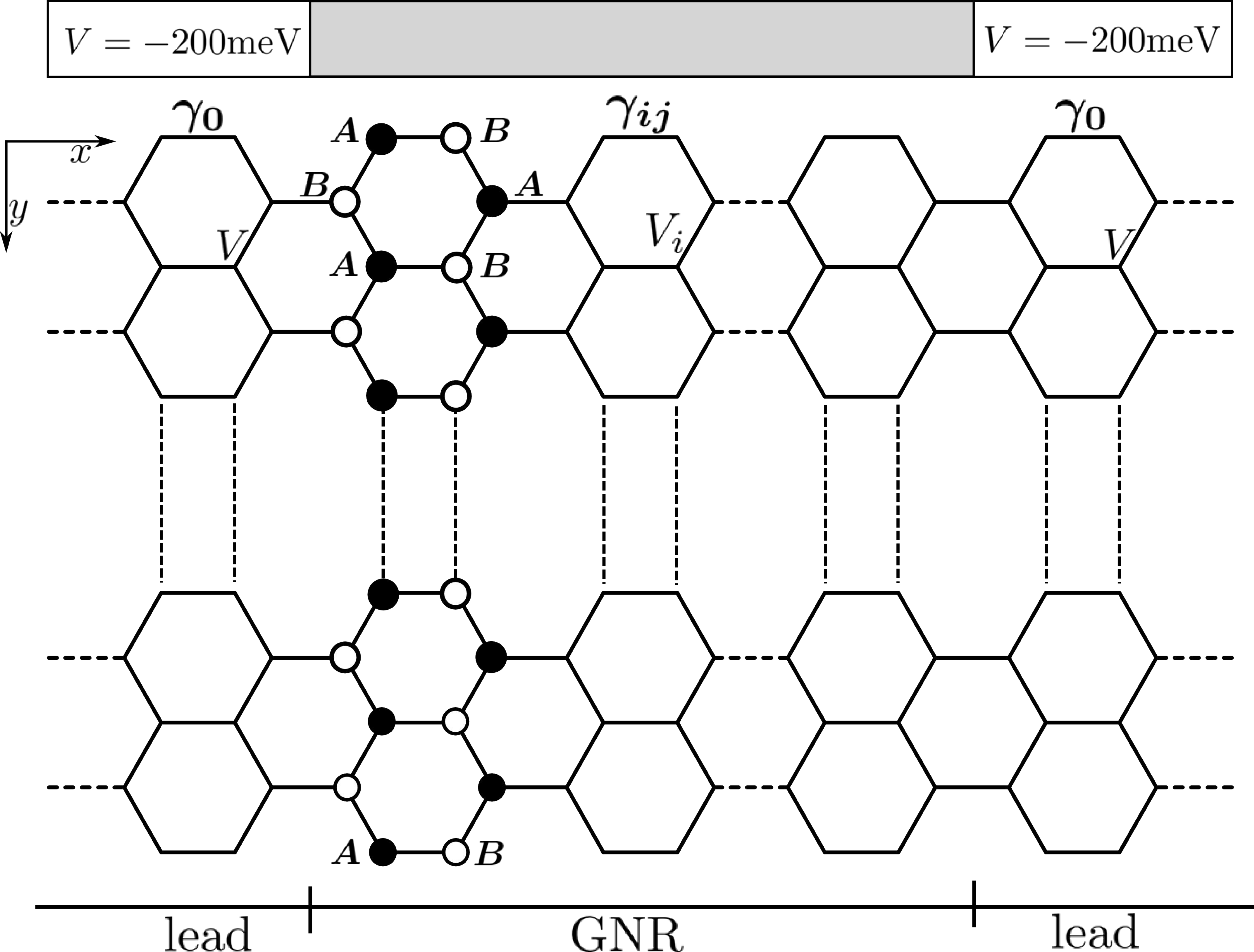}
\caption{\label{fig2} Sketch of the tight-binding model~(\ref{eqn-hamiltonian}) of the GNR junction with armchair boundaries. The system is comprised of two ideal heavily doped leads~($V=-200$meV) and a central suspended region, in which strain modulates the hopping matrix elements $\gamma_{ij}$ and the on-site energy $V_i$.}
\end{figure}

The phase-coherent transport properties of two-terminal devices are encoded in the scattering matrix~\cite{beenakker} which we evaluate using the recursive Green's function technique~\cite{datta,henning1}, applied to the tight-binding model of the GNR sketched in Fig.~\ref{fig2}. Within the Landauer-B{\"u}ttiker formalism~\cite{buttiker} we calculate the device conductance as a function of the Fermi level in the middle part of the GNR, for a given height ($V=-200$meV) of the gate-controlled potential energy step between the doped graphene leads and the suspended part. The resulting device is an $n$-$p$-$n$ ($E_F<0$meV) or $n$-$n'$-$n$ ($E_F>0$meV) graphene junction, where most of the conductance features are determined by scattering from the strain-modified $n$-$p$ or $n$-$n'$ interfaces.

Figure~\ref{fig1}(c) shows the Fermi energy dependence of the zero-temperature conductance of the ribbon with  $L/W=3$, for no strain~(blue) and for artificially imposed homogeneous strain with $w=0.05$~(red). The unstrained armchair GNR is semiconducting; the conductance therefore exhibits a gap around the DP ($E_F=0$meV). The conductance oscillations away from the DP are due to the Fabry-P\'{e}rot-like standing wave resonances in the electron transmission across the potential barrier geometry~\cite{tudorovsky}. For $w=0.05$ homogeneous strain the conductance is completely suppressed for $|E_F|<100$meV. This is because the constant vector potential induced by the homogeneous strain shifts the Dirac cones away from the $K$ and $K'$ corners of the Brillouin zone, perpendicular to the transport direction~\cite{castro_neto}. The misalignment between the Fermi surfaces in the unstrained leads and in the strained suspended region results in a suppression of the conductance in the ballistic regime~\cite{pereira}. The threshold for such an insulating behavior for parameters used in Fig.~\ref{fig1}(c) is $w=0.024$, and can be lowered by reducing the height of the potential step $V$ between the central part of the ribbon and the contacts.

This behavior is significantly changed when the inhomogeneity of the strain is taken into account. Figure~\ref{fig1}(d) shows the Fermi-energy dependence of the conductance for the inhomogeneously strained ribbons shown in Fig.~\ref{fig1}(a). In contrast to Fig.~\ref{fig1}(c), here, we find several groups of additional sharp resonance conductance peaks in the energy range $|E_F|<100$meV. To reveal the nature of each group of these peaks, we analyze the spatial distribution of the corresponding electronic states. Within the Landauer-B{\"u}ttiker formalism, this can be obtained from the response to local perturbations of the scattering amplitudes at the energies close to the resonance conditions~\cite{gasparian}. The reconstructed spatial structure of the resonance states is shown in Fig.~\ref{fig1}(b). As illustrated in the top two rows, the states away from the DP correspond to Fabry-P\'{e}rot-like standing waves formed in the homogeneously strained central part of the structure, to which they are confined due to electron reflection from the interfaces. These resonances appear because the transverse momentum is no longer conserved when the interface region is inhomogeneous, which allows carriers to overcome the misalignment between the Fermi surfaces described above. However, two groups of resonances, in the energy range $-25$meV$<E_F<0$meV just below the DP, display a very different behavior. As shown in the insets in Fig.~\ref{fig1}(d), these resonances occur in almost degenerate pairs, of which we find two groups. The spatial structure of these states, shown in the bottom four rows of Fig.~\ref{fig1}(b), clearly resembles the pseudo-magnetic field distribution.

\begin{figure}[t!]
\includegraphics[width=\columnwidth]{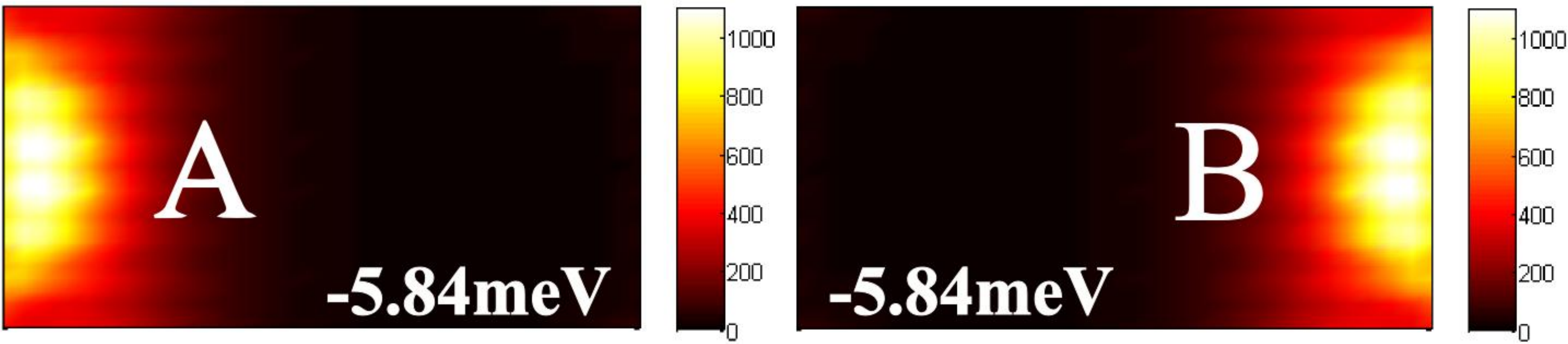}
\caption{\label{fig3} Sublattice-resolved electron amplitude for one of the resonances in Fig.~\ref{fig1}(d)~($L/W=2$, $E=-5.84$meV), obtained by placing the probing perturbation on the $A$~(left panel) or $B$~(right panel) sites.}
\end{figure}

We now demonstrate that this quadruplet of resonances can be attributed to the $n=0$ pseudo-magnetic Landau level. Our main piece of evidence is a unique feature of this LL in graphene, namely that the electron amplitude resides either on the $A$ or $B$ sublattice. The selected sublattice depends on the sign of $\mathcal{B}$ but is independent of the valley~\cite{guinea}. In contrast, higher LLs and Fabry-P\'{e}rot-like resonances occupy both sublattices equally. By selectively placing the probing perturbation on the $A$ or $B$ sites, we find, as illustrated for one example in Fig.~\ref{fig3}, that the low-energy resonances have the following property: their amplitude is high on the $A$-sites near the left end of the ribbon (where $\mathcal{B}<0$), and the amplitude is high on the $B$-site near the right end (where $\mathcal{B}>0$). This fully agrees with the unique feature of the $n=0$ LL described above.

The fact that we find four such low-energy resonances, as well as the dependence of their splitting on the length of the ribbon, further supports this interpretation of the origin of these states. The $y\rightarrow -y$ reflection symmetry maps valleys $K$ and $K'$ onto each other. The resulting symmetric and anti-symmetric superpositions of the two valley manifestations of the $n=0$ LL lead to a splitting of the quadruplet into two groups: a low-energy branch at $E_F \approx-24$meV, which is valley-symmetric and displays a maximum on the symmetry axis, and a high-energy branch at $E_F \approx-7$meV, which displays a nodal line on this axis. Each group splits further into two narrowly spaced lines because of the tunnel coupling of the states $\Psi^0_L$ and $\Psi^0_R$ near the left and right ends of the ribbon, which is provided by the evanescent tails of the electronic wave functions in the middle part, where $\mathcal{B}$ is small. Note that the splitting in each of these pairs is smaller in a longer ribbon, since the overlap of the evanescent tails of $\Psi_L$ and $\Psi_R$ decreases with the separation between the GNR ends.

In conclusion, we describe a unique transport signature of the pseudo-magnetic field in a strained suspended graphene nanoribbon, namely the resonant transmission  via the sublattice-polarized $n=0$ pseudo-magnetic Landau level. These states form near the inhomogeneously strained contact regions and give rise to a characteristic quadruplet of conductance resonances near the Dirac point. The above-proposed analysis is directly applicable to graphene ribbons where high-quality armchair edges are obtained by the oriented growth on patterned SiC substrates~\cite{SiC}, etching of graphene samples with catalytic nanoparticles~\cite{etching}, or chemical derivation~\cite{chemical}. Even though imperfections in the system will lead to the appearance of additional conductance resonances, the $n=0$ Landau level state is protected against the influence of disorder by its unique position at the energy of the Dirac point and its energetic separation from the Fabry-P\'{e}rot-like resonances.

We thank A.~Geim and H.~Ochoa for useful discussions. This project was funded by EC STREP \textit{ConceptGraphene}, EPSRC S\&IA grant, ERC Advanced Grant \textit{Graphene and Beyond}, and by the Royal Society Wolfson Research Merit Award.


\begin{thebibliography}{99}
\bibitem{f_note} Note that we have neglected the possibility of spontaneous wrinkling~\cite{cerda} since strain limits their formation in suspended samples by increasing the transverse rigidity\cite{ochoa}. We have also applied a finite cutoff to regularize the formally divergent displacements~\cite{williams} one finds using linear elasticity theory~\cite{timoshenko} near the corners of clamped ends of the ribbon.
\end{thebibliography}
\end{document}